\documentclass[showpacs,aps,prb,twocolumn,superscriptaddress,10pt]{revtex4-2}

\usepackage{color,xspace}
\usepackage{amsmath, amssymb}
\usepackage{mathrsfs, mathptmx}
\usepackage[colorlinks=true,linkcolor=blue,citecolor=blue,urlcolor=blue]{hyperref}
\usepackage{orcidlink}

\newcommand{\smB}{\ensuremath{\mathbf{B}}\xspace}
\newcommand{\sms}{\ensuremath{\mathbf{S}}\xspace}
\newcommand{\bsigma}{\ensuremath{\boldsymbol{\sigma}}\xspace}
\newcommand{\MnAu}{\ensuremath{\mathrm{Mn}_2\mathrm{Au}}\xspace}

\begin{document}

\title{Revealing the ultra-fast domain wall motion in Manganese Gold through permalloy capping}

\author{Sarah Jenkins}
\affiliation{Faculty of Physics and Center for Nanointegration Duisburg-Essen (CENIDE), University of Duisburg-Essen, 47057 Duisburg, Germany}
\author{Tobias Wagner}
\affiliation{Institute of Physics, Johannes Gutenberg University of Mainz, 55128 Mainz, Germany}
\author{Olena Gomonay}
\affiliation{Institute of Physics, Johannes Gutenberg University of Mainz, 55128 Mainz, Germany}
\author{Karin Everschor-Sitte}
\affiliation{Faculty of Physics and Center for Nanointegration Duisburg-Essen (CENIDE), University of Duisburg-Essen, 47057 Duisburg, Germany}

\date{\today}

\begin{abstract}

Antiferromagnets offer much faster dynamics compared to their ferromagnetic counterparts but their order parameter is extremely difficult to detect and control. So far, controlling the Néel order parameter electrically is limited to only very few materials where Néel spin-orbit torques are allowed by symmetry. In this work, we show that coupling a thin ferromagnet (permalloy) layer on top of an antiferromagnet (Mn$_2$Au) solves a major roadblock -- the controlled reading, writing, and manipulation of antiferromagnetic domains. We confirm by atomistic spin dynamics simulations that the domain wall patterns in the Mn$_2$Au are imprinted on the permalloy, therefore allowing for indirect imaging of the Néel order parameter. Our simulations show that the coupled domain wall structures in Mn$_2$Au-Py bilayers can be manipulated by either acting on the Néel order parameter via Néel spin-orbit torques or by acting on the magnetisation (the ferromagnetic order parameter) via magnetic fields. 
In both cases, we predict ultra-high domain wall speeds on the order of 8.5 km/s. Thus, employing a thin ferromagnetic layer has the potential to easily control the Néel order parameter in antiferromagnets even where Néel spin-orbit torques are forbidden by symmetry. The controlled manipulation of the antiferromagnetic order parameter provides a promising basis for the development of high-density storage and efficient computing technologies working in the THz regime.

\end{abstract}

\maketitle

\section{Introduction}

Spintronics has revolutionised data storage technology with spin current-based technologies such as hard disk drives. However, these devices are reaching the limit of possible technological advances in speed and size due to the intrinsic limits of ferromagnetic (FM) materials. In these devices, antiferromagnetic (AFM) materials are typically only used to provide a preferred direction of magnetisation for the active FM components.  
  Using AFMs instead of FMs as the active component could solve this problem due to their ultra-fast dynamics, lack of external magnetic fields, and temperature stability of the N\'eel ordered state~\cite{MacDonald2011, Gomonay2014, Jungwirth2016, BaltzRevModPhys2018, Jungwirth2018, Jungfleisch2018, Fukami2020}. This brings massive advancements towards ultrafast and ultra high-density spintronics. 

One of the most promising materials for AFM spintronic-based devices is manganese gold (Mn$_2$Au) due to its high Néel temperature, moderate anisotropy, and layered two-sublattice spin structure. It has been predicted that in certain AFMs current-induced spin-orbit-torques (SOTs) can switch the sublattice magnetic orientation~\cite{MacDonald2011, Gomonay2014}. Recently these so-called Néel SOTs have been demonstrated experimentally for CuMnAs~\cite{Wadley2016, Olejnk2017, Olejnk2018, MatallaWagnerPRA2019} and Mn$_2$Au~\cite{MeinertPRApp2018,Behovits2023} and materials with similar AFM ordering~\cite{ChengPRL2020, Hajiri2019, MeinartPRR2020,Shi2020}. 
However, detecting the AFM magnetic signal is far more difficult due to the absence of a net magnetisation.

Recently, a strong exchange coupling between Mn$_2$Au and thin layers of permalloy (Ni$_{80}$Fe$_{20}$) has been observed~\cite{satya,mnau_nife}. As a consequence, the FM domain structure exactly maps the AFM domain structure~\cite{satya, BaltzRevModPhys2018}.
Notably, the coercive field of Mn$_2$Au-Py is an order of magnitude higher ($5000~\mathrm{Oe}$~\cite{satya}) compared to other materials such as CuMnAs ($\sim 200~\mathrm{Oe}$~\cite{Wadley2017}). High coercive fields lead to long-term stability at room temperature. 

Due to the strong coupling, it is anticipated, that the AFM N\'eel vector and the FM magnetisation rotate coherently when an external field is applied to the FM.  
While experimentally the ultrafast dynamics ($\propto~\mathrm{THz}$) and current induced N\'eel SOT switching have yet to be demonstrated in Mn$_2$Au-Py, this report aims to showcase the promising prospects of Mn$_2$Au-Py bilayers.
Atomistic simulations are used to accurately resemble realistic crystal structures and material parameters in dynamics simulations. We derive a  phenomenological model which is in agreement with the simulation results. Besides single domain states of the Mn$_2$Au-Py bilayer system, we simulate the driven dynamics of a domain wall percolating the FM/AFM bilayer by i) an applied magnetic field and ii) Néel SOTs. In both cases, we observe a strongly coupled domain wall motion with domain wall speeds on the order of 8.5 km/s.

\begin{figure}
    \centering
    \includegraphics[width=0.47\textwidth]{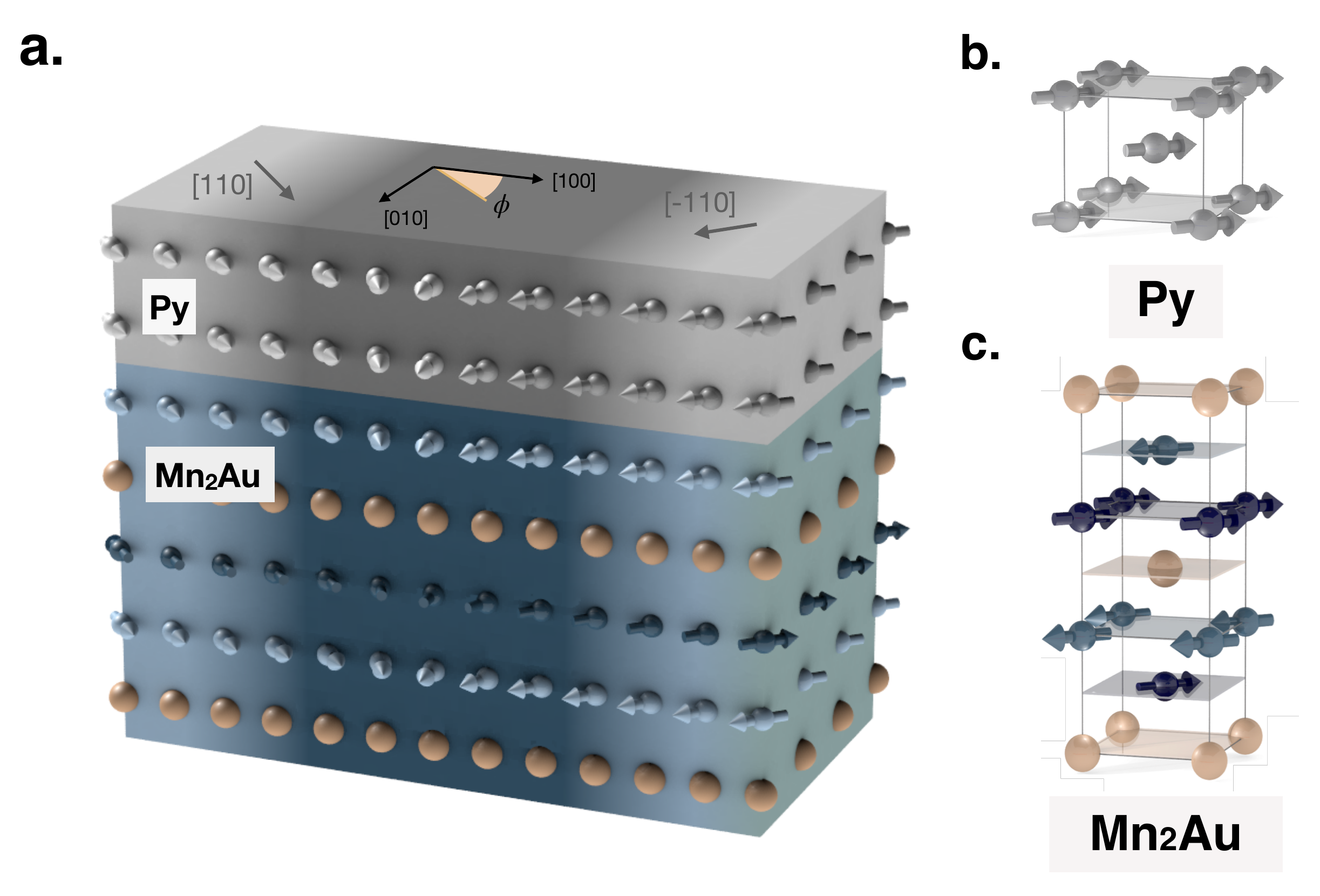}
    \caption{(a) Schematic diagram of a 90-degree magnetic domain wall in a Mn$_2$Au-Py bilayer. The Mn$_2$Au layer is shown in blue and is capped by permalloy shown in grey. 
    (b) and (c) detail the magnetic structures of permalloy and \MnAu, respectively.
    }
    \label{fig:system}
\end{figure}

\section{Model of Mn$_2$Au-Py bilayers}
\label{sec:model}
Here, we model a Mn$_2$Au-Py bilayer system containing a 90-degree domain wall along the [100] direction which is capped by a thin layer of Py. The domain wall from the Mn$_2$Au permeates through to the permalloy due to the strong interface coupling as shown in Fig.~\ref{fig:system} (a). The crystal structures of permalloy and Mn$_2$Au are shown in Fig.~\ref{fig:system}~(b) and~(c) respectively. 
Although py is amorphous in bulk, due to interface lattice matching with the MnAu we presume that Py will have a BCC structure in the thin films we are simulating. The lattice structure of Mn$_2$Au has two AFM coupled planes of Mn atoms and a plane of gold atoms. 
The Hamiltonian $\mathscr{H}$ of the Mn$_2$Au-Py bilayer is comprised of three contributions:
\begin{equation}
    \mathscr{H} = \mathscr{H}_{\mathrm{Mn_2Au}} + \mathscr{H}_{\mathrm{Py}} + \mathscr{H}_{\mathrm{int}},
    \label{eq:H1}
\end{equation}
the Hamiltonian describing Mn$_2$Au ($\mathscr{H}_{\mathrm{Mn_2Au}}$), the Hamiltonian for permalloy ($\mathscr{H}_{\mathrm{Py}}$), and the Hamiltonian for the interface ($\mathscr{H}_{\mathrm{int}}$).

The Hamiltonian for Mn$_2$Au up to the fourth order in anisotropy is given by~\cite{ShickPRB2010}
\begin{multline}
\label{eq:hamiltonianMnAu}
\mathscr{H}_{\mathrm{Mn_2Au}} = 
-\sum_{i<j} \left( \sms_i \mathbf{J}^{\MnAu}_{ij} \sms_j 
+ S_i^z\, \tilde{\mathbf{J}}^{\MnAu}_{ij} S_j^z \right)\\
+\sum_i \left( k_{4} (S_i^z)^4 
+ k_{\phi} (S_i^x)^2 (S_i^y)^2\right)+ \mathscr{H}_{d}, 
\end{multline}
where $\sms_{i,j}$ are unit vectors describing the local spin directions on sites $i,j$. 
The first term describes the exchange energy, which incorporates the second order two-ion anisotropy~\cite{EvansPRBR2020}.
The second term describes the fourth-order anisotropy, and $\mathscr{H}_{d}$ represents the magnetostatic fields. The dominant anisotropy in Mn$_2$Au is the second order easy plane anisotropy. The weaker fourth-order anisotropy terms create easy axis anisotropy directions along the $z$ direction, i.e.\ $k_{4}<0$ as well as a term that breaks the in-plane rotational symmetry favouring the four symmetry equivalent $<110>$ directions, i.e.\ $k_\phi>0$.~\cite{ShickPRB2010} 
For the atomistic spin dynamics, we take three nearest neighbours into account, the first and second nearest neighbour interactions are ferromagnetic and exist between atoms in the same Mn sublattice. The third nearest neighbour interaction is antiferromagnetic between Mn atoms in different sublattices, see App.~\ref{app:simulationdetails} for details.

The Hamiltonian for Py is given by
\begin{equation}
\label{eq:hamiltonianPy}
    \mathscr{H}_{\mathrm{Py}} = -\sum_{i<j}
    \sms_i \mathbf{J}^{\mathrm{Py}}_{ij} \sms_j 
    + \frac{k_c^\mathrm{Py}}{2} \sum_i (S_x^4 + S_y^4 + S_z^4) + \mathscr{H}_{d}.
\end{equation}
The first term describes the exchange energy of the permalloy, the second term is a weak cubic anisotropy.

The interface Hamiltonian
\begin{equation}
\label{eq:hamiltoniani}
    \mathscr{H}_{\mathrm{int}} = -\sum_{i<j} \sms_i \mathbf{J}^{\mathrm{int}}_{ij} \sms_j + \mathscr{H}_{d}
\end{equation}
 acts between the manganese and the permalloy atoms. It leads to a strong coupling between the two layers and imprints the magnetic structures of \MnAu in permalloy. For permalloy~\cite{meilak2019massively} as well as for the interface we use a Heisenberg exchange with a nearest neighbour approximation in the atomistic spin dynamics simulations. 
 In App.~\ref{app:simulationdetails} we list the parameters used for the simulations.

 For investigations with an external field $\mathbf{B}_{\mathrm{ext}}$, we add 
 \begin{equation}
\mathscr{H}_{\mathrm{ext}} =  - \sum_i \mu_s \mathbf{S}_i \cdot \mathbf{B}_{\mathrm{ext}}
 \end{equation}
to the Hamiltonian $\mathscr{H}$ in Eq.~\eqref{eq:H1}.
Note that the external magnetic field is applied to every spin in the system where $\mu_s$ refers to the spin moment in the respective layer ($\mu_{\mathrm{Py}}$ and $\mu_{\mathrm{Mn}}$ are given in Table~\ref{tab:params} in App.~\ref{app:simulationdetails}). 
 As AFMs are largely impervious to magnetic fields, the applied field mostly affects the permalloy. However, due to the strong exchange coupling between the FM and the AFM, the magnetic order in Mn$_2$Au is indirectly affected by an applied field as well.

 \section{Static properties of single domain states in the bilayer system}
 \label{sec:sdstatics}
We performed Monte Carlo simulations on the pure Mn$_2$Au system (for a system size of 8 nm $\times$ 8 nm $\times$ 5 nm) which reproduced the low-temperature ground state spin structure with the magnetisation of the two Mn sublattices 180 degrees apart, see Fig.~\ref{fig:system} (c). The easy axes are along the $<$110$>$  directions, as expected from the rotational anisotropy and in agreement with previous neutron scattering experiments~\cite{Barthem2013} and theoretical calculations~\cite{ShickPRB2010, ostler}.

\begin{figure}[tb]
    \centering
    \includegraphics[width=0.5\textwidth]{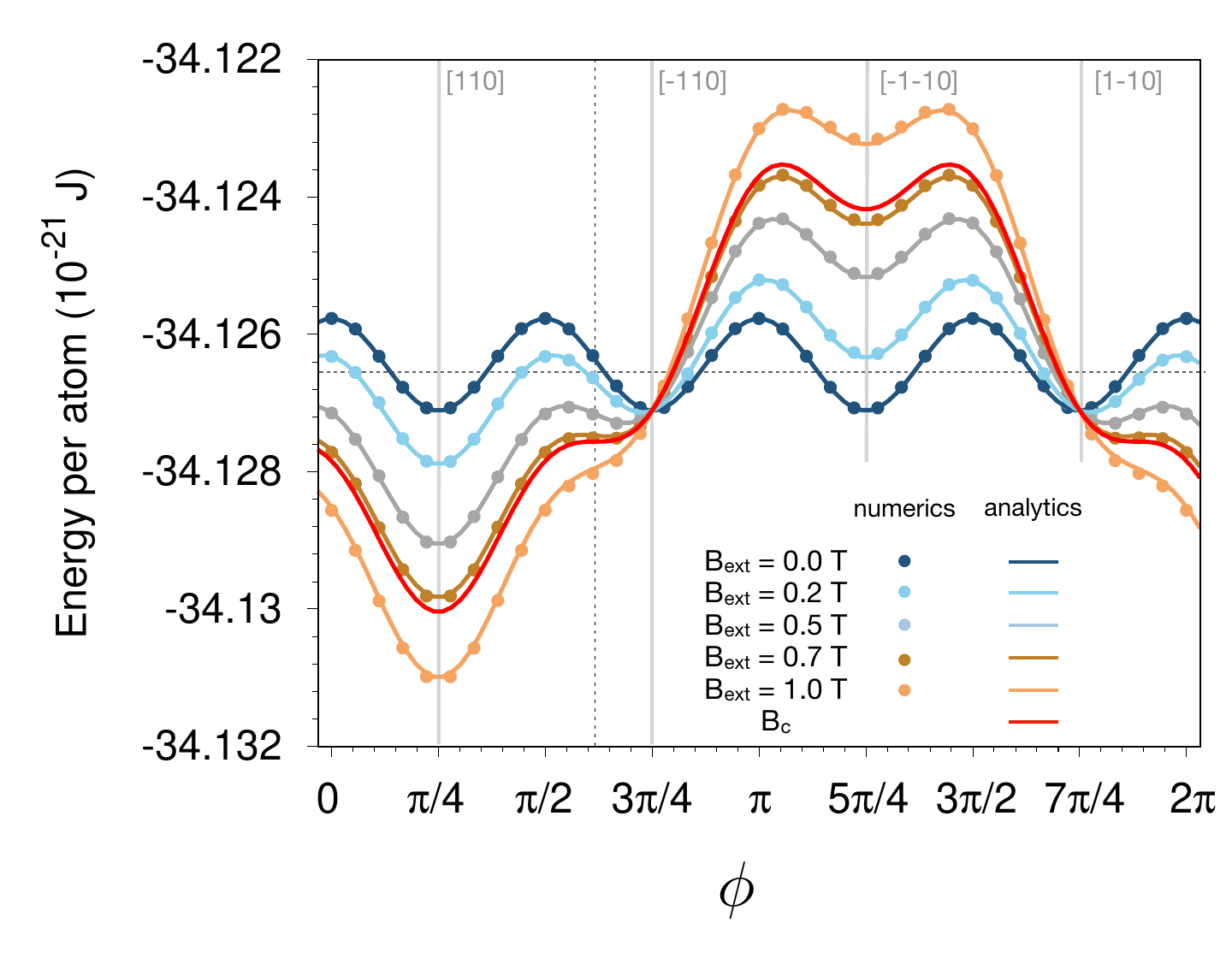}
    \caption{Energy surfaces of an \MnAu-Py bilayer (average energy per atom) with an applied field of varying strength along the [110] direction, with the critical field $B_c\approx 0.7555 T$. The MnAu thickness was 5 nm and the permalloy thickness was 1 nm.
     }
    \label{fig:cmcfield}
\end{figure}

For the bilayer system (of lateral size 8 nm $\times$ 8 nm, and thickness 5 nm of Mn$_2$Au and 1nm of permalloy) we find that the ground state of Mn$_2$Au is imprinted into the Py as expected due to the strong interlayer exchange coupling~\cite{satya}. Performing constrained Monte Carlo simulations~\cite{Asselin2010ConstrainedAnisotropy} we computed the energy surfaces for a single domain state in the bilayer system~\footnote{
We constrained the direction of the magnetisation of a single Mn sublattice while allowing all other spins to relax to their equilibrium spin structures.}.  
When a magnetic field is applied the four-fold degeneracy in the energy surface is broken. The applied magnetic field was along the $[110]$ direction, and the energy surface, therefore, favours this easy axis over the other three easy axis directions.
To characterise the in-plane spin direction, we introduce the angle $\phi$ as the angle from the $[100]$ axis in the $xy$ plane, see Fig.~\ref{fig:system}. 
By scanning all angles $\phi$ we obtain the energy surfaces for varying field strengths as shown in Fig.~\ref{fig:cmcfield}. As expected, the field direction, i.e.\ $\phi=\pi/4$, is energetically preferred. For magnetic domains oriented perpendicular to the field direction, i.e.\ $\phi= 3\pi/4$ and $\phi=7\pi/4$, the energy of the system is independent of the strength of the magnetic field. 
The easy axis direction which is anti-parallel to the magnetic field, i.e.\ $\phi=5\pi/4$, gets enhanced in energy with increasing field strength.
A spherical version of Fig.~\ref{fig:cmcfield} can be found in App.~\ref{app:singledomain}.

The strong in-plane anisotropy of Mn$_2$Au and the strong interlayer coupling causes the spins in the two layers to be effectively in-plane. For a single domain state with an applied magnetic field along the $[110]$ direction (as shown in Fig.~\ref{fig:cmcfield}), the physics reduces to an effective Hamiltonian $\mathscr{H}_{\mathrm{bilayer}}$ described by the angle $\phi$:
 \begin{equation}
\mathscr{H}_{\mathrm{bilayer}}(\phi) = E_0 
+ c_{\phi}  k_\phi \cos 4\phi
- c_B  \mu_{\mathrm{Py}} B_\textrm{ext} \cos(\phi - \pi/4).
\label{eq:analyticsB}
 \end{equation}
Here $E_0$ is the energy of the system without the four-fold in-plane anisotropy term. The magnetic field couples to the magnetisation, i.e.\ the average spin moment of permalloy per atom $\mu_{\mathrm{Py}}$, in the form of a $\cos\phi + \sin\phi$ term which can be rewritten as a shifted cosine function, as $\cos\phi + \sin\phi= \sqrt{2} \cos(\phi - \pi/4)$. The constants $c_\phi$ and $c_B$ depend on the system. $c_\phi$ is proportional to the layer thickness of the \MnAu, whereas $c_B$ is proportional to the layer thickness of the permalloy.

Upon increasing the magnetic field strength above a critical field value $B_c$, the metastable states (previously oriented approximately perpendicular to the applied field) disappear. At the critical field value $B_c$, 
the local minima close to $\phi=3\pi/4$ and $7\pi/4$ turn into saddle points, with 
i) $\partial \mathscr{H}_{\mathrm{bilayer}}(\phi)/ \partial \phi|_{\phi=\tilde{\phi}} =0$, ii) $\partial^2 \mathscr{H}_{\mathrm{bilayer}}(\phi)/ \partial \phi^2|_{\phi=\tilde{\phi}} =0$ and iii) $\partial^3 \mathscr{H}_{\mathrm{bilayer}}(\phi)/ \partial \phi^3|_{\phi=\tilde{\phi}} =0$.
Solving these two equations close to $\phi =3\pi/4$ we obtain
\begin{equation}
    B_c =\frac{16 \sqrt{6} k_\phi c_\phi}{9 \mu_{\mathrm{Py}} c_B} 
\end{equation}
with $\tilde{\phi} = 3\pi/4 - \delta\phi\approx 1.9357$ being independent of the model parameters.  
Using the parameters summarized in App.~\ref{app:simulationdetails}) we obtain from the simulation results $E_0\approx -34.1264\,\cdot 10^{-21} \mathrm{J}/\mathrm{atom}$, $c_\phi=0.8375$ and $c_B=0.2603$.
For the critical field value, we obtain $B_c\approx 0.7555 T$.

\section{Domain walls in Mn$_2$Au-Py bilayer systems}
\label{sec:dwstatics}
\begin{figure}[t]
    \centering
    \includegraphics[width=0.4\textwidth]{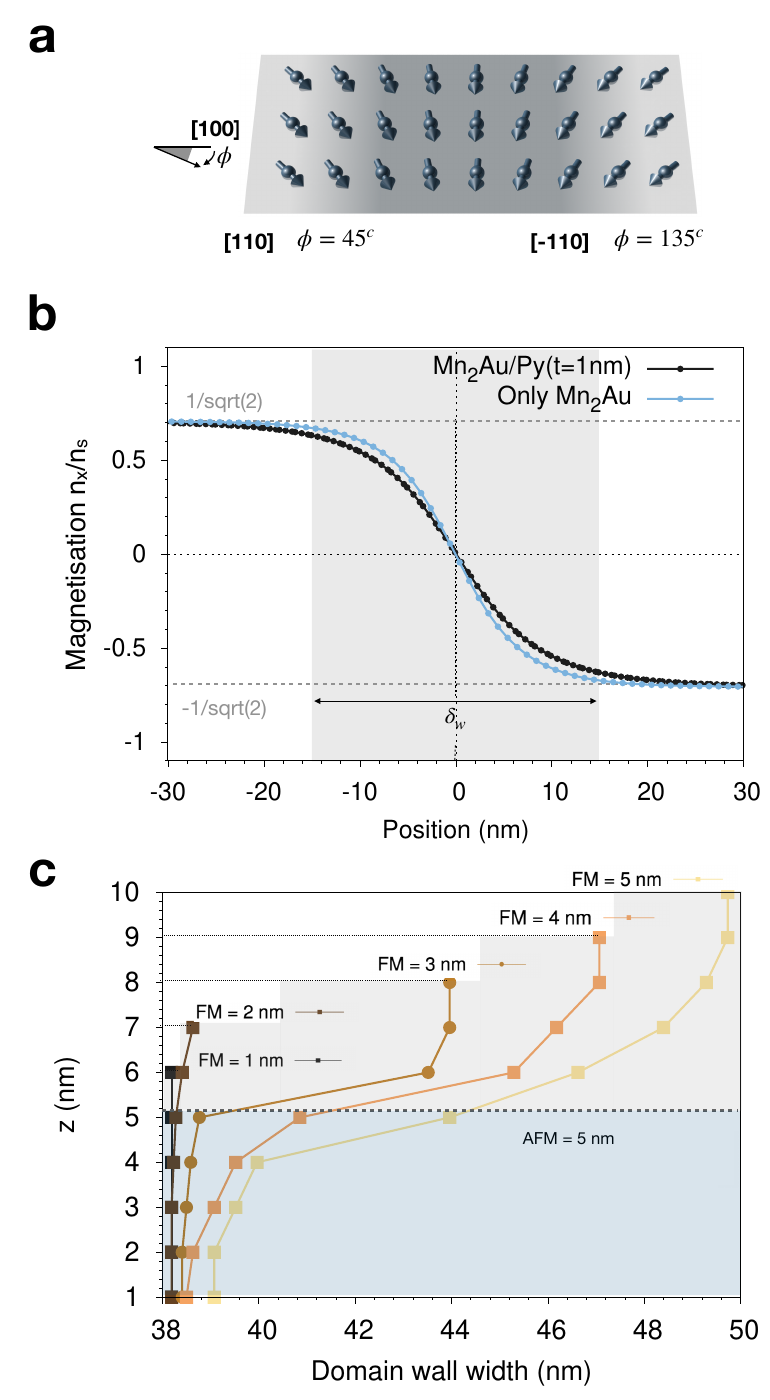}
    \caption{Static domain wall structures in \MnAu-Py bilayers.
    a) Sketch of a 90-degree domain wall. 
    b) Domain wall profiles $n_x$ for a 5 nm thick MnAu layer with a 1 nm permalloy capping and without permalloy capping. 
    c) Domain wall widths along the $z$ direction for varying permalloy capping layer thicknesses.
    }
    \label{fig:dwwidths}
\end{figure}

In this section, we study a static domain wall across a ribbon bilayer system of dimensions 1000 nm along $x$ and 40 nm along $y$. The $z$ height of MnAu is kept constant at 5 nm, the py height is initialy 1 nm, but this is later varied.  
A 90-degree domain wall from the [110] direction to the [1-10] direction was initialized at $x$ = 100 nm in both the \MnAu and the Py, see Fig.~\ref{fig:system} and Fig.~\ref{fig:dwwidths}a). Our key findings are summarised below:
i) The domain wall of \MnAu is imprinted on the permalloy capping.
ii) The bilayer structure has a net remaining magnetisation from the permalloy which then allows the observation of domain walls in the AFM via the capping~\cite{satya}.
iii) The permalloy capping increases the average domain wall width of the bilayer system compared to pure \MnAu.
iv) The domain wall width varies along the vertical direction, with increasing domain wall width towards the top (i.e.\ FM) surface.

We computed the domain wall profile of the bilayer using a Monte Carlo integrator. The simulation was run for one million Monte Carlo steps at 0 K until the magnetic structure remained stationary. Fig.~\ref{fig:dwwidths}a) shows the normalized (height averaged) $x$-component of the magnetic order parameter along the ribbon for the bilayer system and a comparison to pure \MnAu. The magnetic order parameter used to model the domain wall profile is the magnetisation for the permalloy and the magnetisation of one sublattice for the \MnAu. We denote this normalized (height averaged) magnetic order parameter by $\mathbf{n}$.
The domain wall width was extracted by fitting the $x$-component, $n_x=\cos\Theta(x)$, to the expression of the profile function 
\begin{equation}
\Theta(x)=\arctan \left[ \exp\left(\frac{ x-x_0}{ \delta_{\mathrm{dw}}}\right)\right] +\frac{\pi}{4}.
\end{equation}
Here $x_0$ is the position of the domain wall, and $\delta_{\mathrm{dw}}$ the domain wall width.
For the 90-degree domain wall, $n_x$ varies from $n_x = 1/\sqrt{2}$ on the left, over to $n_x = 0$ in the center of the domain wall to $n_x = -1/\sqrt{2}$ on the right end, as shown in Fig.~\ref{fig:dwwidths}a) and b).

Our numerical results reveal a domain wall width of $\delta_{\mathrm{dw}}\approx29.3$~nm for the case of pure \MnAu.
As the natural domain wall width of permalloy is about an order of magnitude larger, the average domain wall width of the bilayer system depends on the relative thickness of the permalloy to \MnAu.
For 1~nm permalloy on top of 5~nm \MnAu, see Fig.~\ref{fig:dwwidths}, we obtain a domain wall width of $\delta_{\mathrm{dw}}\approx38.2$~nm, i.e.\ an increase of about $30\%$. 
For such a thin permalloy capping we do not observe a significant change of the domain wall width along the $z$ direction.

Fig.~\ref{fig:dwwidths}c) displays the change of the domain wall width with height ($z$-direction) for the device for varying capping layer thicknesses of permalloy.
We find that, as expected, i) the domain wall becomes wider towards the FM and ii) there is a larger domain wall width change for thicker permalloy capping. The change in the domain wall width comes from the competition between the magnetocrystaline anisotropy and the exchange interactions, as the exchange interactions and anisotropy vary from bulk MnAu to bulk Py across the interface the domain wall widths vary. For thin Py capping layers the domain wall width in the Py matches the MnAu due to the strong interface exchange coupling. 

\section{Manipulating domain walls in Mn$_2$Au-Py bilayer systems}
In this section, we focus on the dynamics of the previously considered domain walls in Mn$_2$Au-Py bilayer systems.
Two external driving mechanisms are investigated separately to move the domain wall: an external magnetic field and a current, which induces N\'eel SOTs. 

The system dynamics were calculated by solving the stochastic Landau-Lifshitz-Gilbert equation~\cite{Garcia-Palacios1998Langevin-dynamicsParticles}:
\begin{equation}
\frac{\partial \mathbf{S}_i}{\partial t} = -\frac{\gamma_e }{1 + \lambda^2} \left[ \mathbf{S}_i \times \mathbf{B}_{\mathrm{eff}} + \lambda \mathbf{S}_i \times \left(\mathbf{S}_i \times  \mathbf{B}_{\mathrm{eff}} \right) \right],
\label{eq:LLG}
\end{equation}
which  models the interaction of an atomic spin moment $\mathbf{S}_i$ with an effective field $\mathbf{B}_{\text{eff}}=- \delta \mathscr{H}/\delta \mathbf{S}_i$.
The effective field causes the atomic moments to precess around the field, where the frequency of precession is determined by the gyromagnetic ratio of an electron ($\gamma_e =1.76 \times 10^{11}$ rad s$^{-1}$T$^{-1}$). The damping constant $\lambda$ is defined in each of the two materials, see Table~\ref{tab:params} in App.~\ref{app:simulationdetails}.
For our numerical results, Eq.~\eqref{eq:LLG} was solved using a Heun scheme~\cite{vampire}.

In the following, we considered a 1~nm permalloy capping on top of 5~nm \MnAu.
The lateral dimensions of our system remain the same as in Sec.~\ref{sec:dwstatics}, i.e.\ 1000 nm along $x$, 40 nm along $y$. Our starting point is a relaxed 90-degree domain wall in the bilayer system from the [110] direction to the [1-10] located at $x$ = 100 nm.

\subsection{Magnetic field-driven domain wall dynamics}

To study the magnetic field-driven domain wall dynamics, we apply a magnetic field along the [110] direction, making this direction the energetically preferred one. To reduce the energy of the bilayer system, the [110] domain in the permalloy is shifted via a domain wall motion. Due to the strong interface exchange coupling between the two layers, the domain wall in the \MnAu is dragged along by the permalloy, leading to a coupled domain wall motion.

\begin{figure*}[t]
    \centering
    \includegraphics[width=0.9\textwidth]{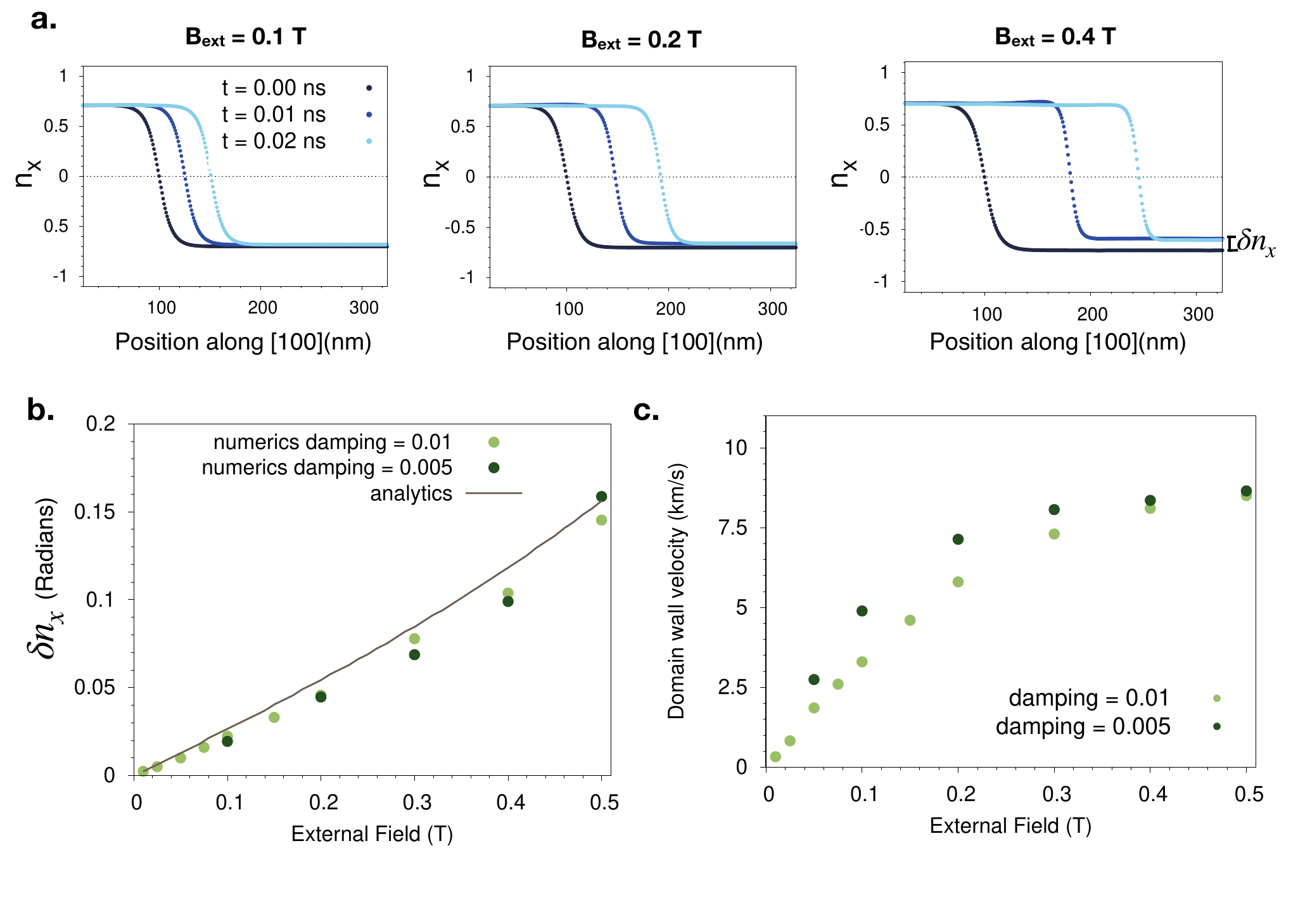}
    \caption{Magnetic field-driven domain wall motion for a Py thickness of 1 nm; a) domain wall profile snapshots for varying field strengths. This is a height-averaged magnetisation of one AFM sublattice. 
    b) Change in domain orientation $\delta n_x$ as a function of magnetic field strength. 
    c) domain wall velocity as a function of applied field strength. 
    }
    \label{fig:field}
\end{figure*}

We analyse this coupled domain wall motion for different magnetic field strengths. In Fig.~\ref{fig:field} a) we show snapshots of the domain wall profile at different times for various magnetic field strengths. Besides extending the [110] domain, the field rotates the [1-10] domain such that the $x$ component of the magnetic order parameter assumes the shifted value of $n_x= -1/\sqrt{2} + \delta n_x$. 
In Fig.~\ref{fig:field} b) we display $\delta n_x$ as a function of the applied field strength. At the critical field $B_c$, the domain wall flips to a uniform state, which induces a jump in $\delta n_x$.
This is in agreement with Sec.~\ref{sec:sdstatics}, where we find a magnetic field induced shift $\delta \phi$ in the position of the energy minimum for the domain that is oriented approximately perpendicular to the magnetic field, with $\delta n_x=1/\sqrt{2}- \cos(\delta \phi-\pi/2)$.

In Fig.~\ref{fig:field} c) we show the domain wall velocities as a function of the applied magnetic field strength. As expected, the domain wall moves faster for stronger applied fields. 
The existence of a critical magnetic field above which the domain wall is flipped into a single domain state, however, puts a boundary on the maximally possible domain wall speed induced by magnetic fields. 
In our numerics, the magnetic field-driven domain walls have velocities between 0.1 - 8.5 km/s with maximum possible values of 8.5 km/s close to $B_c$. 

\subsection{N\'eel spin-orbit torque driven domain wall dynamics}

In the \MnAu-Py bilayers the SOTs are induced by the Au atoms. The Au atoms cause spin-dependent scattering in the \MnAu and due to the spin symmetry around the Au sites this leads to a N\'eel SOT acting in opposite directions on the different sublattices. Therefore, the SOTs act only on the Mn atoms and there are no SOTs in the Py capping layer. 

To study the domain wall dynamics induced by N\'eel SOTs, we add the N\'eel SOT $\smB_{\mathrm{SOT}}$ to the effective field in the LLG equation, Eq.~\eqref{eq:LLG}
\begin{equation}
    \mathbf{B}_{\text{eff}} = - \frac{\delta \mathscr{H}}{\delta \mathbf{S}_i} + \mathbf{B}_{\text{SOT}}
\end{equation}
where
\begin{equation}
\label{eq:SOT}
\smB_{\mathrm{SOT}} = B_{\mathrm{NT}}^{\mathrm{SOT}} \left(\bsigma - \lambda \mathbf{S} \times \bsigma\right) + B_{\mathrm{FT}}^{\mathrm{SOT}} \left(\mathbf{S}\times \bsigma + \lambda \bsigma \right),
\end{equation}
 comprises both the field-like ($B_{\mathrm{FT}}^{\mathrm{SOT}}$) and damping like N\'eel SOTs ($B_{\mathrm{NT}}^{\mathrm{SOT}}$)~\cite{Manchon2019b, Meo_2023} acting on the \MnAu layers. The SOTs act in opposite directions in each Mn sublattice due to the spin splitting off the Au atoms. The directions of the SOTs are defined by the spin polarisation unit vector $\bsigma$. The spin polarisation unit vector acts perpendicular to the electron flow. Here the electron flow is along [001], and the SOTs acts along [110] and 
[-1-10] in the two Mn sublattices respectively, effectively shifting the domain wall along the [100] axis.
The SOT field strengths $B_{\mathrm{FT}}^{\mathrm{SOT}}$ and $B_{\mathrm{NT}}^{\mathrm{SOT}}$ for an atomistic monolayer are given by~\cite{Manchon2019b}
\begin{subequations}
    \begin{align}
    B_{\mathrm{FT}}^{\mathrm{SOT}} &= \frac{\hbar j_e \theta_\mathrm{SH} a^2}{2e \mu_{\mathrm{Mn}}} \\
    B_{\mathrm{NT}}^{\mathrm{SOT}} &=     \beta_{\mathrm{SOT}} B_{\mathrm{FT}}      
\end{align}
\end{subequations}
where $j_e$ is the injected current density and $a$ is the unit cell size. The spin Hall angle  $\theta_\mathrm{SH}$ gives the conversion efficiency of electrical current into spin current.
$\beta_{\mathrm{SOT}}$ is an empirical scaling factor that relates the strength of the precessional term with the relaxation term, here assumed to equal one.

\begin{figure*}[t]
    \centering
    \includegraphics[width=0.9\textwidth]{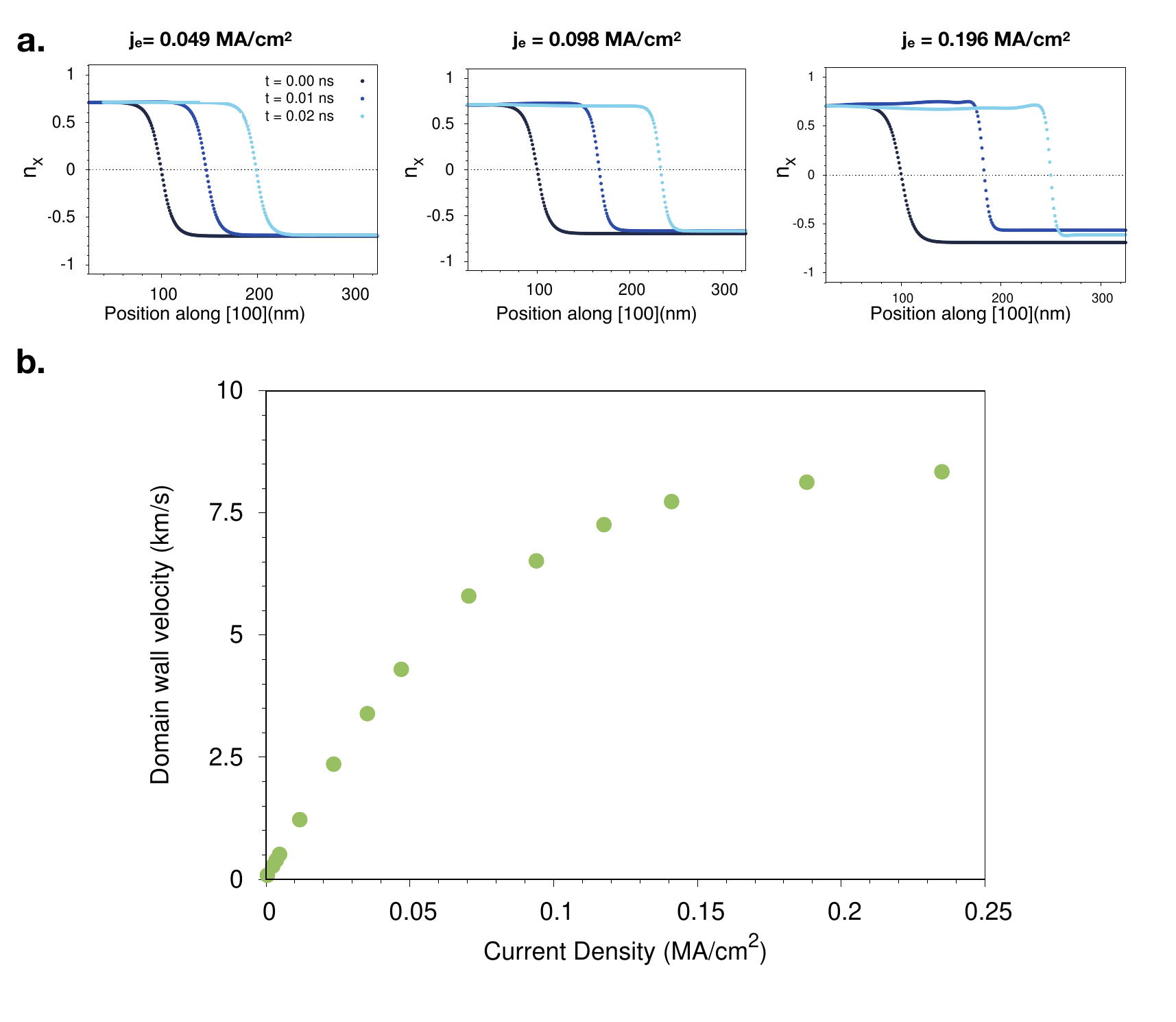}
    \caption{Current driven domain wall motion for a Py thickness of 1 nm; a) domain wall profile snapshots for varying currents. Domain wall profile snapshots for varying current strengths. This is a height-averaged magnetisation of one AFM sublattice. 
    b) domain wall velocity as a function of applied field strength.}
    \label{fig:sot}
\end{figure*}

In Fig. \ref{fig:sot}a we show the domain wall profiles at 0.00 ns, 0.01 ns, and 0.02 ns for low to higher spin current strengths; $j_e= 0.049\,  \mathrm{MA}/\mathrm{cm}^2$, $0.098\,\mathrm{MA}/\mathrm{cm}^2$ and $0.196\,  \mathrm{MA}/\mathrm{cm}^2$ corresponding to $B_{\mathrm{NT}}^{\mathrm{SOT}}= 0.01 T$, 0.02 and 0.04. As expected, for higher spin current strengths the current causes deformations in the shape of the domain wall profile over time. 
After the initial transient dynamics due to switching on the current, the domain wall reaches a steady state shape and motion in which we compute the domain wall velocity as the average of the gradient of position vs.\ time.
For our chosen parameters we find the maximum velocity of Néel SOT-driven domain walls to be again close to 8.5 km/s even for low currents on the order of 1 MA/cm$^2$.

\section{Discussion and Summary}
We have shown that bilayer systems consisting of a thin permalloy capping on top of \MnAu display strongly coupled domain wall structures, where the domain wall structure of the AFM is imprinted on the FM. A clear advantage of the thin permalloy capping is that the magnetic dynamics of \MnAu become experimentally observable via the magnetisation of the permalloy \cite{satya}.
Furthermore, we have shown that the coupled magnetic textures can be manipulated by two mechanisms: i) magnetic fields and ii) N\'eel SOTs.
While the applied magnetic field acts mainly in the permalloy, the N\'eel SOTs act in the \MnAu. In both cases, the strong coupling between the layers causes the spin dynamics in one layer to affect the spin dynamics in the other layer which finally induces a collective domain wall motion. Notably the domain wall motion is independent of the driving mechanism and the material it originated.
For both, magnetic field and N\'eel SOT-driven domain walls we obtain velocities on the order of 8.5 km/s.
Such a controlled manipulation of the AFM
is a key requirement for AFM-based high-density storage and efficient THz information and computing technologies.

\section{Acknowledgements}
We thank Richard Evans, Ricardo~Rama-Eiroa, Rub\'en M.~Otxoa, Roy Chantrell, Martin Jourdan, and Satya Prakash Bommanaboyena for fruitful discussions. We acknowledge funding from the German Research Foundation (DFG) - TRR 173 – 268565370 Spin + X: spin in its collective environment (project B12 and A05), Project No. 320163632 (Emmy Noether), 
CRC/TRR 270 - 405553726
This project made use of the \textsc{Magnitude} cluster, a high-performance computing facility provided by the University of Duisburg-Essen. 
\appendix

\section{Simulation Details}
\label{app:simulationdetails}
The parameters used for the atomistic simulations are summarised in Table~\ref{tab:params} and explained in more detail below.

\begin{table}
\centering 
\begin{ruledtabular}
\begin{tabular}{c c c c c }
Interaction & $J_{xx}$ & $J_{yy}$ & $J_{zz}$ & Unit \\
\hline 
$\mathbf{J}^{\MnAu}_{1}$ & -1.094296 & -1.094296 & -1.086911 & $10^{-20}$ J (per link) \\
$\mathbf{J}^{\MnAu}_{2}$ & -1.469234 & -1.469234 & -1.459319 & $10^{-20}$ J (per link)  \\
$\mathbf{J}^{\MnAu}_{3}$ &  0.318261 &  0.318261 &  0.318261 & $10^{-20}$ J (per link)  \\
$\tilde{\mathbf{J}}^{\MnAu}_{1}$ & 0.0 & 0.0 & 0.007385 & $10^{-20}$ J (per link)  \\
$\tilde{\mathbf{J}}^{\MnAu}_{2}$ & 0.0 & 0.0 & 0.009915& $10^{-20}$ J (per link)  \\
$\tilde{\mathbf{J}}^{\MnAu}_{3}$ &  0.0 &  0.0 &  0.318261 & $10^{-20}$ J (per link) \\
$\tilde{\mathbf{J}}^{Py}$ & 0.009915 & 0.009915 & 0.009915& $10^{-20}$ J (per link) \\
$\tilde{\mathbf{J}}^{int}$ &  0.318261 &  0.318261 &  0.318261 & $10^{-20}$ J (per link)  \\
\hline 
 & Parameter & \multicolumn{2}{c}{Value} & Unit \\
\hline 
&$a,b$           & \multicolumn{2}{c}{3.327}                         & \AA \\
&$c$             & \multicolumn{2}{c}{8.539}                     & \AA \\
&$\mu_\mathrm{Mn}$ 
& \multicolumn{2}{c}{4.0}                     & $\mu_{\mathrm{B}}$ \\
& $\mu_\mathrm{Py}$
& \multicolumn{2}{c}{1.6}                     & $\mu_{\mathrm{B}}$ \\

&$k_4^{\MnAu}$           & \multicolumn{2}{c}{$-1.60218 \times 10^{-24}$}     & J/atom \\
&$k^{\MnAu}_{\phi}$        & \multicolumn{2}{c}{$8.00109 \times 10^{-25}$}     & J/atom \\
&$k_c^{\mathrm{Py}}$           & \multicolumn{2}{c}{$1.0 \times 10^{-26}$}     & J/atom \\
&$\lambda_{\MnAu}$       & \multicolumn{2}{c}{0.01}                          & $-$ \\
&$\lambda_{\mathrm{Py}}$       & \multicolumn{2}{c}{0.01}                          & $-$ \\
\end{tabular}
\end{ruledtabular}
\caption{Model parameters for \MnAu used in the simulations. Isotropic exchange parameters are taken from Khmelevskyi \textit{et al.}~\cite{KhmelevskyiAPL2008} and magnetic anisotropies are taken from Shick \textit{et al.}~\cite{ShickPRB2010}. The magnetic moment of the Mn sites is taken from experimental measurements of Barthem~\textit{et al.}~\cite{Barthem2013}.
}
\label{tab:params}
\end{table}

\begin{figure}
\centering
\includegraphics[width= 0.57\columnwidth]{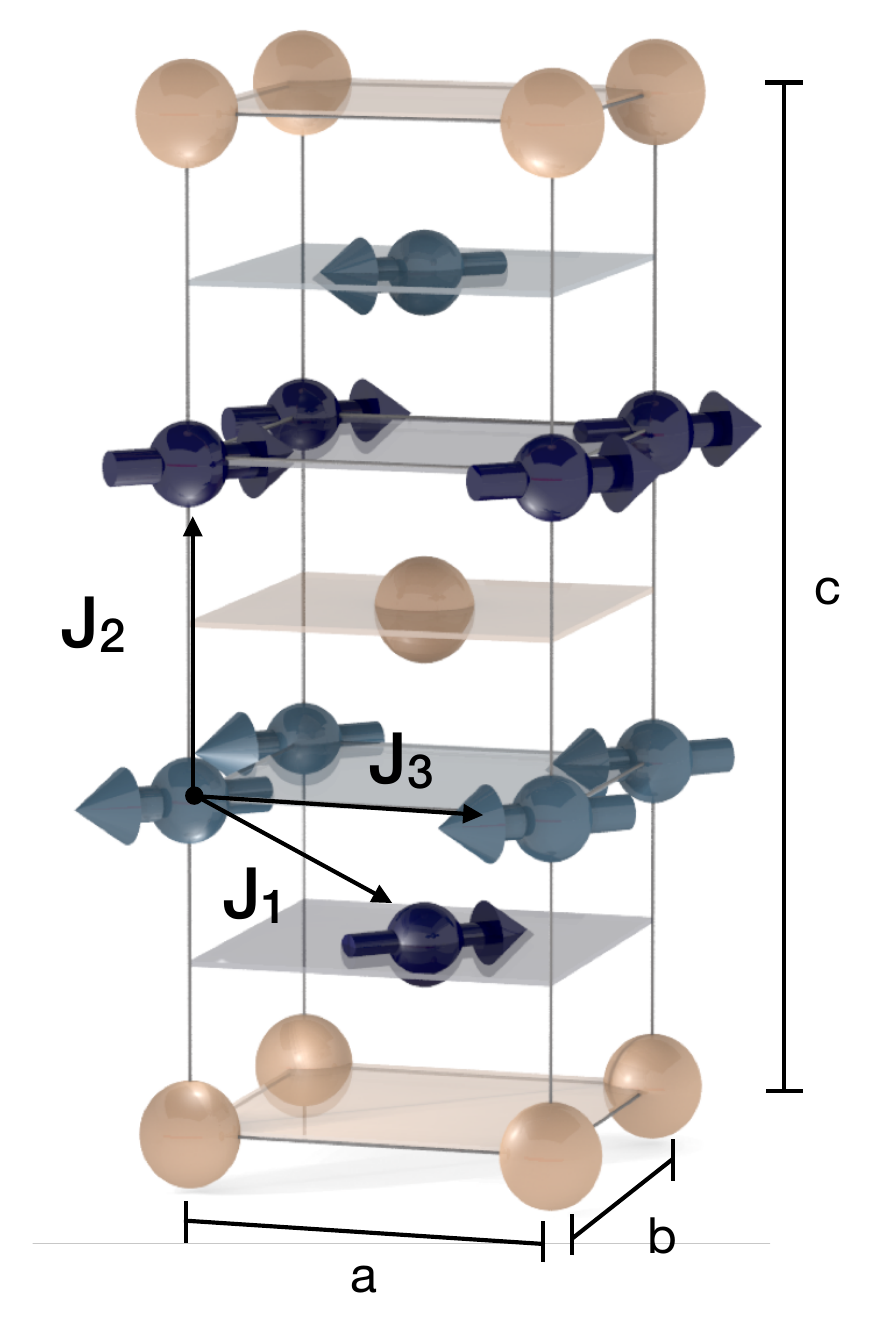}
\caption{Crystal structure of \MnAu including the nearest, next, and next-next nearest neighbour interactions.}
\label{fig:neighbours}
\end{figure}

The crystal structure of \MnAu with unit cell dimensions $a,b$ and $c$ is shown in Fig.~\ref{fig:neighbours}.
$J_1$, $J_2$ and $J_3$ indicate the isotropic exchange parameters for the nearest neighbour, next nearest neighbour, and next-next nearest neighbour, at distances 2.180~\AA, 2.853~\AA, and 3.327~\AA, respectively.

In the simulations for the isotropic exchange in \MnAu, the exchange interactions were taken into account up to the next-next nearest neighbour with parameters from Khmelevskyi \textit{et al.}~\cite{KhmelevskyiAPL2008}. 
The magnetic anisotropies are taken from Shick \textit{et al.}~\cite{ShickPRB2010} and the magnetic moment of the manganese sites is taken from experimental measurements of Barthem \textit{et al.}~\cite{Barthem2013}. 

For permalloy, we consider only the nearest neighbour FM interaction and the weak cubic anisotropy with $k_c^{\mathrm{Py}}$ taken from Ellis \textit{et al.}~\cite{permalloy}.

For the interface between \MnAu and permalloy, we chose an exchange constant of $J^{\mathrm{int}} = 0.31 \times 10^{-20}$ being roughly 20\% of the bulk exchange value of \MnAu. This value is motivated by recent experiments where the interface coupling has been found to be strong~\cite{satya}. For the damping parameters $\lambda_\mathrm{Py}$ and $\lambda_\mathrm{\MnAu}$ we used 0.01 
if not stated otherwise.

\section{Single Domain states / spherical version}
\label{app:singledomain}
\begin{figure}
\centering
\includegraphics[width= 0.45\textwidth]{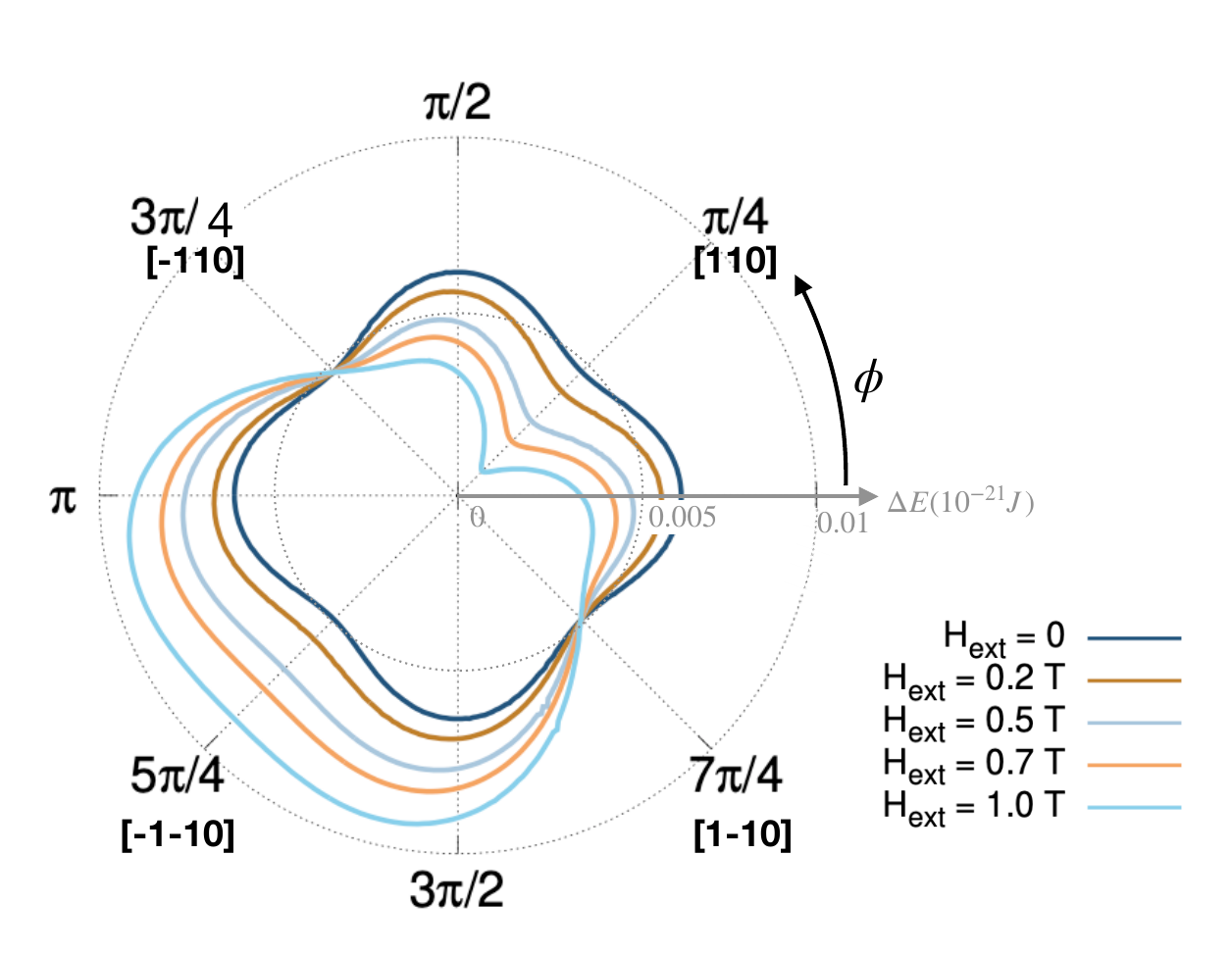}
\caption{Spherical plot of energy surfaces of an \MnAu-Py bilayer (average energy per atom) with an applied field of varying strength along the [110] direction. Larger radii correspond to larger energies. }
\label{fig:interfacecircle}
\end{figure}

Fig.~\ref{fig:interfacecircle} shows the spherical version of Fig.~\ref{fig:cmcfield} of the main text. It reveals the transition from the four-fold rotational symmetry in the absence of a magnetic field, to a remaining mirror symmetry in the direction of the magnetic field.

\bibliography{library}

\end{document}